\documentclass[oldversion]{aa}  
\usepackage{graphicx}
\usepackage{txfonts}
\usepackage{longtable}
\usepackage{natbib}
\usepackage{ulem}
\bibpunct{(}{)}{;}{a}{}{,}

\begin{document}
\title{The spectroscopically confirmed huge cosmic structure at $z=0.55$}
\subtitle{}
\titlerunning{The huge cosmic structure at $z=0.55$}
\authorrunning{Tanaka et al.}

\author{M. Tanaka\inst{1}, A. Finoguenov\inst{2,3}, T. Kodama\inst{4}, Y. Koyama\inst{5},  B. Maughan\inst{6}, F. Nakata\inst{7}	}


\institute{European Southern Observatory, Karl-Schwarzschild-Str. 2, D-85748 Garching bei M\"{u}nchen, Germany
	\email{mtanaka@eso.org}
	\and
 	Max-Planck-Institut f\"{u}r extraterrestrische Physik, Giessenbachstrasse, D-85748 Garching bei M\"{u}nchen, Germany
	\and
	University of Maryland, Baltimore County, 1000 Hilltop Circle,  Baltimore, MD 21250, USA
	\and
        National Astronomical Observatory of Japan, Mitaka, Tokyo 181-8588, Japan
	\and
	Department of Astronomy, School of Science, University of Tokyo, Tokyo 113--0033, Japan
	\and
	HH Wills Physics Laboratory, Tyndall Avenue,  Bristol, BS8 1TL, UK
	\and
        Subaru Telescope, National Astronomical Observatory of Japan, 650 North A'ohoku Place, Hilo, HI 96720, USA
}

\date{Received; accepted }

\abstract{
We report on the spectroscopic confirmation of a huge cosmic structure
around the CL0016 cluster at $z=0.55$.
We made wide-field imaging observations of the surrounding regions
of the cluster and identified more than 30 concentrations of red galaxies
near the cluster redshift.  The follow-up spectroscopic observations
of the most prominent part of the structure confirmed 14 systems
close to the cluster redshift, roughly half of which have a positive
probability of being bound to the cluster dynamically.
We also made an X-ray follow-up, which detected extended X-ray emissions from
70\% of the systems in the X-ray surveyed region.
The observed structure is among the richest 
ever observed in the distant Universe.
It will be an ideal site for quantifying environmental variations in the
galaxy properties and effects of large-scale structure on galaxy evolution.
}{}{}{}{}
 


\maketitle

\section{Introduction}

The cosmic large-scale structure has grown from the nearly uniform matter
distribution in the early Universe to the rich filamentary and clumpy
structure that we observe locally today \citep{york00,colless01}.
Recent large redshift surveys with 8m telescopes unveiled
structure up to $z\sim1$ \citep{scoville07}.
Cosmic structure is now probed at even higher redshifts by
narrow-band surveys (e.g., \citealt{shimasaku03,matsuda05}).
Galaxies are arrayed in filamentary structure with a typical scale of 10 Mpc
\citep{colberg05} and filaments form the complicated cosmic web.
Galaxy clusters often lie at nodes of filaments, and
a prominent structure is often expected around them \citep{bond96,springel05}
and so is observed (e.g.,  \citealt{gal08,tanaka08}).
Blank field surveys tend to miss a very rich structure because
it is rare, but pointed observations of rich clusters can
discover it.
In this paper, we present an extremely rich structure around
the CL0016+16 cluster at $z=0.55$.

The structure around the cluster was first noted by \citet{koo81}.
Several follow-up observations have been carried out since then
\citep{hughes95,munn97,hughes98,dressler99}.
In \citet{tanaka07a} we presented the spectroscopic confirmation
of the prominent structure around the cluster in the form of
two clumpy filaments. 
The structure appeared to extend outside of the observed field.
Motivated by this, we made further imaging and spectroscopic observations of
the surrounding regions to follow the entire structure.

The layout of the paper is as follows.
In Sect. 2, we summarize our imaging and spectroscopic observations
and move on to present the huge structure in Sect. 3.
The paper is summarized in Sect. 4.
Unless otherwise stated, we adopt H$_0=70\rm km\ s^{-1}\ Mpc^{-1}$,
$\Omega_{\rm M}=0.3$, and $\Omega_\Lambda =0.7$.

\section{Observations}

\subsection{Optical imaging observations}

The imaging observations of the surrounding regions of the CL0016 cluster
were carried out with Suprime-Cam \citep{miyazaki02} in $V$ and $i$ bands between
August 2007 and January 2008.
Six pointings were observed in total (NE, NW, E, W, SW, and SE of the central
field presented in \citealt{tanaka07a}). 
The observing conditions were good and all the images were smoothed to
a common FWHM size of $0.7\arcsec$.
The exposure times varied from pointing to pointing, but they were
typically 18 min in $V$ and 6 min in $i$.
Data were reduced using the generic imaging pipeline \citep{yagi02}
with custom designed scripts.
The photometric zero-points were obtained using the standard stars observed
in the same nights.
The Galactic extinction was corrected using the dust map by \citet{schlegel98}.

\begin{figure}
\centering
\includegraphics[width=9.2cm]{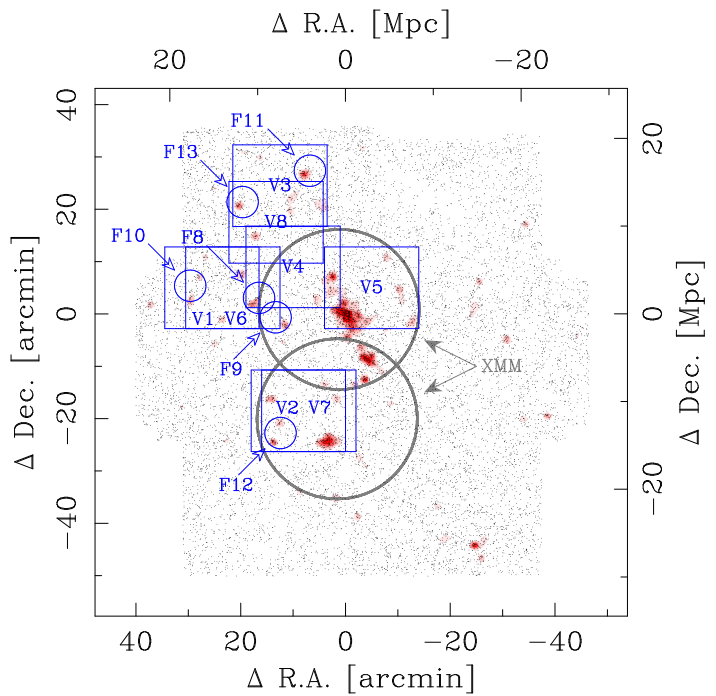}
\caption{
The dots show $V-i$ selected red galaxies
($\Delta|V-i|<0.2$ from the cluster red sequence)
and the shades show the density of the selected galaxies.
The two large circles are the regions probed by XMM.
The observed FOCAS and VIMOS fields are indicated with the small circles
and squares, respectively.
The top and right axes show comoving scales.
}
\label{fig:lss}
\end{figure}

Red galaxies around the cluster redshift were selected using the $V-i$ color
($\Delta |V-i|<0.2$ from the cluster red sequence).
The distribution of the selected galaxies is shown in Fig. \ref{fig:lss}.
The extremely rich CL0016 cluster is at ($\Delta$R.A.,$\Delta$Dec.)=($0'$,$0'$).
A clumpy structure extends to the south reaching to another rich
cluster at ($\Delta$R.A.,$\Delta$Dec.)=($3'$,$-27'$) \citep{hughes98}.
There are more than 10 concentrations of red galaxies in the E-NE of the cluster.
There are also several scattered groups to the west of the cluster.
The prominent structure seems to extend more than 20 Mpc, and
it is potentially among the richest ever discovered
in the distant Universe.

\subsection{X-ray Observations}

CL0016 was originally observed by XMM-Newton for a 38ks observation
on 2000 Dec. 29 (OBSID 0111000101 and 0111000201; \citealt{worrall03}).
This was then followed by a 62ks observation on
2007 Dec. 14 (OBSID 0502860101) that was offset
by $\sim20$ arcminutes to the south to cover the large-scale structure
detected in our optical survey \citep{kodama05}.
In addition to the standard data processing of the EPIC data, which was done
using XMMSAS version 6.5
\citep{saxton05},
we performed a more conservative removal of time intervals affected
by solar flares \citep{zhang04}.
To increase our capability of detecting extended, low surface
brightness features, we applied the quadruple background subtraction
\citep{finoguenov07} and also checked for high background,
identifying a hot MOS2 chip
(CCD id=5)  in the OBSID 0502860101. We removed the 
identified hot chip from any further analysis.
The resulting cleaned exposure time for the on-source observation
amounts to 35ksec (pn, m1, m2) and off source  $\sim45$ ksec 
(m1 and m2) and 30 ksec (pn).

After the background was estimated for each observation and each
instrument separately, we produced the final mosaic of cleaned images.
We used the prescription of \citet{finoguenov09} for extended source
detection, which consists of removal of the PSF model for each detected point
source from the data before applying the extended source search algorithm.
We found 28 extended X-ray sources in the 0.35 square degrees of the XMM 
mosaic, 7 of which are associated with the target cluster and groups at
the cluster redshift.

For clumps 4 and 6 (see below for the definitions of the clumps),
the X-ray data were sufficient for spectral analysis.
Spectra were extracted with an aperture of radius 1.5\arcmin\ for each camera,
with background spectra extracted from 4 regions of the same size at the same off-axis angle.
This ensured that the vignetting of the X-ray photons
was similar in the source and background regions.
The data from the different cameras
were fit simultaneously with an absorbed APEC model \citep{smith01}
with the absorbing column fixed at
the Galactic value \citep{dickey90}.
The metal abundance was left free, but was not tightly constrained by
the data for either clump.
We measured temperatures for clump 4 and 6 as 
$3.4\pm0.4\ \rm keV$ (consistent with \citealt{worrall03})
and $4.0\pm0.9\ \rm keV$, respectively.

\subsection{Optical spectroscopic follow-up observations}

Following the imaging observations, we carried out
spectroscopic observations to probe the most prominent part of
the structure using FOCAS on Subaru 
and VIMOS on VLT Melipal. 
The FOCAS observations were taken on 2008 Aug. 9-10 in visitor mode.
The 300B grism blazed at $5500\AA$ with the SY47 order sorting filter
was used giving a resolving power of $R\sim500$.
The observing conditions were very good and the seeing was $\sim0.7$ arcsec.
Six fields were observed in total (F8-F13).
Integration times were 60 min (15min $\times$ 4 shots) for all of the fields,
except for F13, which was observed 45 min in total.
The data were reduced using a custom-designed pipeline. 


The VIMOS observations were taken between October and December 2008 in queue mode.
The MR grism with GG495 order sorting filter was used, providing
a resolving power of $R\sim500$.
The observing conditions were variable -- thin to clear conditions with
$\lesssim1.2\arcsec$ seeing.  Some of the observations
were executed under relatively
bright conditions.  It should be noted that not all the planned observations
were executed, and the resulting field coverage
is somewhat patchy.
Each mask was exposed for 16 min $\times 4$ shots $=$ 64 min.
The data were reduced in the same way as for the FOCAS data.


We visually inspected all the reduced spectra and assigned
redshifts and confidence flags.
We also collected redshifts from the literature
\citep{hughes95,hughes98,munn97,dressler99,tanaka07a}.
The final spectroscopic catalog contains 1202 secure redshifts
and 132 possible redshifts including 48 duplicated measurements.
The median and dispersion of the differences between the duplicated redshifts
are $<0.0001$ and 0.0007, respectively.
There is no sign of systematic offset between the different observing runs.
We regard redshifts with quality flag $\geq3$ from \citet{munn97}
and flag $\leq3$ from \citet{dressler99} as secure.
In this paper, we use only objects with secure redshifts.

\section{The cosmic large-Scale structure at $z=0.55$}

\begin{figure*}
\centering
\includegraphics[width=11.9cm]{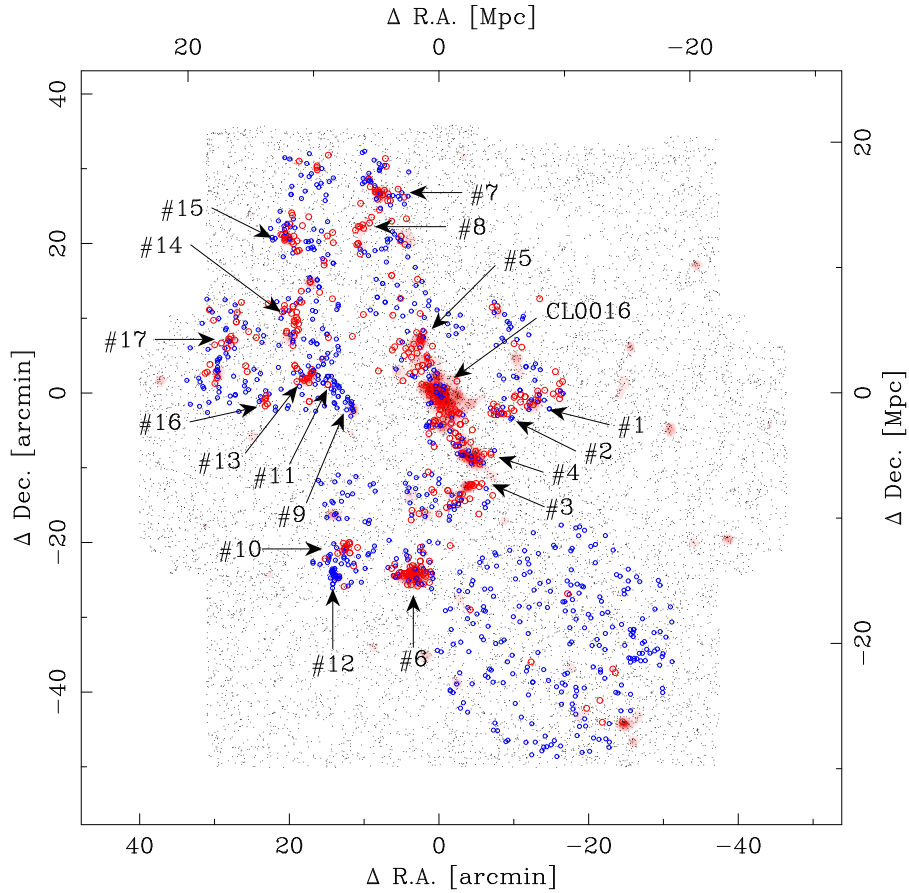}
\caption{
The distribution of the spectroscopic objects.
The red circles show galaxies at $0.530<z<0.565$,
which includes most CL0016 cluster members in the redshift space.
The blue circles show
those outside of the redshift range.
Apparent concentrations of  galaxies are indicated with the arrows.
The dots and shades are the same as in Fig. \ref{fig:lss}.
}
\label{fig:lss_spec}
\end{figure*}

We present in Fig \ref{fig:lss_spec} the spectroscopically confirmed
large-scale structure around the CL0016 cluster.
We confirm many of the photometrically identified groups of galaxies with
a few exceptions, which turn out to be background systems.
Apparent concentrations of spectroscopic galaxies are numbered in the figure
and their properties are summarized in Table \ref{tab:clump_props}.
Their coordinates are given in J2000 in the 2nd and 3rd columns.
Redshifts (4th column) and velocity dispersions (6th column) are
estimated with the $2\sigma$-clipped biweight estimator and gapper method
\citep{beers90}, respectively, using galaxies within 0.5 Mpc (physical) from
the clump centers.
The numbers of galaxies used after the clipping are listed in the 5th column.
The centers are defined as the positions of the outstandingly bright
spectroscopic members or the averaged centers of a few brightest members.
The errors were estimated by bootstrapping the input objects taking
the individual redshift errors into account.
The 7th and 8th columns are masses ($\rm M_{200}$ in units of $10^{13}\ \rm M_\odot$)
from the velocity dispersion \citep{carlberg97} and X-ray, respectively.

The statistics are still poor, and it is in some cases not obvious
whether a clump is a bound system.
We classify a clump as a group if there is a strong spatial and redshift
concentration of galaxies based on a visual inspection.
The clumps with weak spatial concentrations may be loose groups,
but we are not yet sure at this point.
Several statistical methods of quantifying filamentary structure
are proposed in the literature (e.g., minimum spanning tree; \citealt{barrow85}),
but we cannot apply these methods to our data due to
the very irregular spatial sampling of the spectroscopic galaxies.

The prominent structure extends towards the NE of the cluster, where we confirm
7 systems (clumps 7, 8, 13, 14, 15, 16, and 17).
There seems to be two distinct structures there.
One is at $z=0.556$, only a slight distance from the CL0016 cluster
($\Delta z\sim+0.01$),
consisting of clumps 8, 14, and 16.
The other one is at the cluster redshift formed by the other clumps.
The main structure extends towards west and SE as reported in \citet{tanaka07a},
but there are several concentrations of red galaxies,
for which we did not perform spectroscopic follow-up observations.
We did not explore S-SW of the cluster further beyond clump 6,
but it is interesting to note that the spectroscopically observed galaxy at
($\Delta$R.A, $\Delta$.Dec.)$\sim$($-25'$,$-45'$) by \citet{munn97}
is an outstandingly bright galaxy at the center of a concentration of red galaxies.
This likely cD galaxy of a possible group implies
that the structure extends farther.
It will not be surprising if the structure extends beyond the probed field.

We quantified the dynamical status of the structure based on the classical Newtonian dynamics
\citep{beers82,hughes95,lubin98}.
We estimated a probability that each system is dynamically bound to the CL0016 cluster.
The entire tidal field around the cluster should ideally be taken into account,
but in our particular case, the CL0016 cluster is by far the most massive system
in the field, and it likely dominates the tidal field.
Here we simply solve the two-body problem.
Most of the masses from the velocity dispersions have large
errors due to poor statistics.
We used X-ray masses where available.
For the CL0016 cluster, we used the hydrostatic mass derived by \citet{solovyeva07}.
For clumps 4 and 6, we used our derived temperatures to estimate masses from
the mass-temperature relation of Sanderson et al. (2003;  we used their relation D
and included self-similar evolution). For the other clumps detected in X-ray,
masses were estimated using the $L_X-M$ relation from \citet{leauthaud09}.
The luminosity derived masses of clumps 4 and 6 are consistent with
those derived from the temperature.

The probabilities are summarized in the 9th column in Table \ref{tab:clump_props}.
The probabilities for clumps 1, 2, 3, and 6 are somewhat lower compared to
those by \citet{tanaka07a} due to the lower mass of the CL0016 cluster
adopted here. (\citealt{tanaka07a} took the mass from velocity dispersion)
Many of the confirmed groups close to the cluster are likely to be bound to the cluster,
suggesting they would fall into the cluster in the future.
Clumps 8, 14, and 16 are not bound to the cluster, and
the NE structure at $z=0.556$ is not within the gravitational reach of the cluster.
However, given its close proximity ($\Delta z\sim0.01$) to the cluster,
the structure is likely spatially connected to the cluster, being
a part of the huge cosmic structure around CL0016.
The rest of the clumps in the NE may be bound to the cluster
albeit with the large uncertainties.

This is the first confirmation of such a prominent clumpy
structure in the distant Universe, which is among the richest ever observed.
The structure extends well over 20 Mpc, and some of the clumps are probably
bound to the central cluster.
This suggests that it is a physically connected structure.
However, the structure can be further extended beyond the probed field
given that some of the spectroscopically confirmed groups are
located near the edge of the field.
More observations are needed to observe
the gigantic cosmic structure in its entirety.

\begin{table*}
\caption{
Properties of the clumps.
}
\label{tab:clump_props}
\centering
\tiny
\hspace*{-0.7cm}
\begin{tabular}{llllrcllll}
\hline\hline
       & R.A.     & Dec.            & $z$         & $N_{spec}$ &   $\sigma$  [$\rm km s^{-1}$] & $\rm M_{200, opt}$ & $\rm M_{200, X-ray}$  & $P_{bound}$ & comment\\
\hline\vspace{2pt}
CL0016	& 00 18 33.4 & 16 26 14 	& $0.5471_{-0.0027}^{+0.0032}$ 	& 24 	& $1683_{-380}^{+131}$ 	& $612.39_{-328.34}^{+155.08}$ 	& $119\pm2$		& --- 	 		 & $M_{200}$ is from \citet{solovyeva07}\\\vspace{0.5pt}
Clump1	& 00 17 40.2 & 16 24 59 	& $0.5474_{-0.0003}^{+0.0004}$ 	&  6 	& $128_{-89}^{+ 42}$ 	& $0.27_{-0.26}^{+0.36}$ 	& $<5.1$ ($2\sigma$)	& $94^{+0}_{-63}$ 	 & filament/group reported in \citet{tanaka07a}\\\vspace{0.5pt}
Clump2	& 00 17 58.9 & 16 23 28 	& $0.5512_{-0.0008}^{+0.0007}$ 	&  8 	& $248_{-120}^{+ 53}$ 	& $1.95_{-1.68}^{+1.53}$ 	& $<3.5$ ($2\sigma$)	& $44^{+35}_{-33}$ 	 & filament/group reported in \citet{tanaka07a}\\\vspace{0.5pt}
Clump3	& 00 18 16.4 & 16 14 02 	& $0.5482_{-0.0024}^{+0.0020}$ 	&  6 	& $550_{-441}^{+174}$ 	& $21.36_{-21.20}^{+27.50}$ 	& $6.8\pm0.6$		& $81^{+6}_{-63}$ 	 & group reported in \citet{tanaka07a}\\\vspace{0.5pt}
Clump4	& 00 18 17.4 & 16 17 39 	& $0.5493_{-0.0016}^{+0.0015}$ 	& 12 	& $862_{-315}^{+117}$ 	& $82.10_{-61.15}^{+38.38}$ 	& $17\pm4$		& $70^{+19}_{-43}$ 	 & cluster reported in \citet{hughes95}\\\vspace{0.5pt}
Clump5	& 00 18 43.7 & 16 33 17 	& $0.5509_{-0.0018}^{+0.0012}$ 	&  9 	& $576_{-285}^{+136}$ 	& $24.44_{-21.29}^{+21.69}$ 	& $7.3\pm1.1$		& $52^{+32}_{-35}$ 	 & group\\\vspace{0.5pt}
Clump6	& 00 18 47.6 & 16 02 14 	& $0.5409_{-0.0010}^{+0.0007}$ 	& 21 	& $562_{-92}^{+203}$ 	& $22.85_{-9.53}^{+34.71}$ 	& $22\pm9$		& $ 0^{+35}_{-0}$ 	 & cluster reported in \citet{hughes98}\\\vspace{0.5pt}
Clump7	& 00 19 06.5 & 16 53 02 	& $0.5447_{-0.0005}^{+0.0007}$ 	&  8 	& $258_{-113}^{+301}$ 	& $2.21_{-1.81}^{+20.21}$ 	& ---			& $43^{+36}_{-43}$ 	 & group\\\vspace{0.5pt}
Clump8	& 00 19 17.6 & 16 48 29 	& $0.5568_{-0.0007}^{+0.0005}$ 	&  4 	& $119_{-86}^{+ 60}$ 	& $0.21_{-0.21}^{+0.52}$ 	& ---			& $ 0^{+0}_{-0}$ 	 & group?\\\vspace{0.5pt}
Clump9	& 00 19 22.2 & 16 24 32 	& $0.7278_{-0.0013}^{+0.0019}$ 	&  5 	& $519_{-512}^{+  0}$ 	& $16.10_{-16.10}^{+0.02}$ 	& ---			& $ 0^{+0}_{-0}$ 	 & background loose group?\\\vspace{0.5pt}
Clump10	& 00 19 25.7 & 16 05 23 	& $0.5427_{-0.0002}^{+0.0004}$ 	&  8 	& $ 75_{-17}^{+148}$ 	& $0.05_{-0.03}^{+1.36}$ 	& $<4.9$ ($2\sigma$)	& $10^{+53}_{-10}$ 	 & group\\\vspace{0.5pt}
Clump11	& 00 19 31.5 & 16 27 44 	& $0.6253_{-0.0006}^{+0.0001}$ 	&  4 	& $ 14_{ -2}^{+122}$ 	& $0.00_{-0.00}^{+0.31}$ 	& $7.8\pm1.5$		& $ 0^{+0}_{-0}$ 	 & background group\\\vspace{0.5pt}
Clump12	& 00 19 32.4 & 16 02 03 	& $0.6264_{-0.0007}^{+0.0003}$ 	& 13 	& $199_{-98}^{+173}$ 	& $0.96_{-0.83}^{+5.33}$ 	& $8.9\pm1.8$		& $ 0^{+0}_{-0}$ 	 & background group\\\vspace{0.5pt}
Clump13	& 00 19 47.9 & 16 28 11 	& $0.5445_{-0.0007}^{+0.0006}$ 	& 10 	& $594_{-491}^{+ 80}$ 	& $26.96_{-26.82}^{+12.35}$ 	& ---			& $55^{+30}_{-43}$ 	 & group\\\vspace{0.5pt}
Clump14	& 00 19 53.3 & 16 36 02 	& $0.5584_{-0.0046}^{+0.0009}$ 	&  6 	& $246_{-109}^{+981}$ 	& $1.90_{-1.57}^{+234.02}$ 	& ---			& $ 0^{+0}_{-0}$ 	 & loose group?\\\vspace{0.5pt}
Clump15	& 00 19 58.4 & 16 46 58 	& $0.5425_{-0.0006}^{+0.0003}$ 	&  9 	& $128_{-48}^{+319}$ 	& $0.27_{-0.20}^{+11.23}$ 	& ---			& $ 3^{+53}_{-3}$ 	 & group\\\vspace{0.5pt}
Clump16	& 00 20 10.4 & 16 25 21 	& $0.5550_{-0.0003}^{+0.0004}$ 	&  4 	& $ 64_{-45}^{+ 53}$ 	& $0.03_{-0.03}^{+0.17}$ 	& ---			& $ 0^{+3}_{-0}$ 	 & group?\\\vspace{0.5pt}
Clump17	& 00 20 29.9 & 16 33 23 	& $0.5443_{-0.0008}^{+0.0010}$ 	&  7 	& $324_{-179}^{+ 78}$ 	& $4.37_{-3.98}^{+4.00}$ 	& ---			& $34^{+43}_{-34}$ 	 & group\\

\hline
\end{tabular}
\end{table*}

\section{Summary}

We have reported the spectroscopic confirmation of the huge structure at $z=0.55$.
We carried out the wide-field imaging and spectroscopic observations
of the rich cluster CL0016 at $z=0.55$.
We spectroscopically confirmed many of the photometrically identified galaxy groups 
at the cluster redshift.
The structure probed will be among the most prominent 
confirmed structures in the distant Universe,
although the structure likely extends farther.

The huge structure provides us with a unique opportunity to
quantify the dependence of galaxy properties
on the whole range of environment at $z=0.55$.
In particular, we can address the effects of
large-scale environments on galaxy evolution.
The amplitude of clustering of groups around the CL0016 is unusually high.
This enables us to effectively extend our previous studies to
large physical separations,
where the role of structure formation on galaxy properties has not been revealed yet.
We will report on differences in spectral properties of galaxies
in different environments, together with more detailed X-ray analysis,
in a forthcoming paper.

\begin{acknowledgements}
This study is based on data collected at the Subaru Telescope,
which is operated by the National Astronomical Observatory of Japan,
through program S07B-SV148 and S08B-025.
We thank Dr. Furusawa and Dr. Hattori for their help during the observations.
This study is also based on observations obtained at the ESO
Very Large Telescope
through program 082.A-0201 and 
on observations obtained with XMM-Newton, an ESA science mission with instruments
and contributions directly funded by ESA Member States and NASA.
We thank S. Finnegan and J. Leyland for their preliminary analysis of the XMM data.
This work was financially supported in part by the Grant-in-Aid for Scientific Research
(No. 18684004 and 21340045) by the Japanese Ministry of Education,
Culture, Sports, and Science.
YK acknowledges support from the Japan Society for the Promotion of
Science (JSPS) through JSPS fellowships for Young Scientists.
AF has been partially supported trough NASA grant NNX08AD93G to UMBC.
We thank the anonymous referee for helpful comments, which improved the paper.
\end{acknowledgements}
\vspace{-0.8cm}

\bibliographystyle{aa}
\bibliography{12929ref}

\end{document}